\documentclass[prl,twocolumn,showpacs,preprintnumbers,amsmath,amssymb,floatfix,reprint]{revtex4}
\usepackage{hyperref}
\usepackage{epsfig}
\usepackage{color}
\begin{document}

\title{High-pressure lubricity at the meso- and nanoscale}

\author{A. Vanossi$^{1,2}$, A. Benassi$^3$, N. Varini$^{4,5}$, and E. Tosatti$^{2,1,6}$}
\affiliation{
$^1$ CNR-IOM Democritos National Simulation Center, Via Bonomea 265, 34136 Trieste, Italy \\
$^2$ International School for Advanced Studies (SISSA), Via Bonomea 265, 34136 Trieste, Italy \\
$^3$ Empa, Materials Science and Technology, \"Uberlandstrasse 129, 8600 D\"ubendorf, Switzerland \\
$^4$ Research \& Development, Curtin University, GPO Box U 1987, Perth, Western Australia 6845 \\
$^5$ iVEC, 26 Dick Perry Ave, Kensington WA 6151, Australia \\
$^6$ International Centre for Theoretical Physics (ICTP), Strada Costiera 11, 34014 Trieste, Italy
}

\begin{abstract}
The increase of sliding friction upon increasing load is a classic in the macroscopic world.
Here we discuss the possibility that friction rise might sometimes turn into a drop when,
at the mesoscale and nanoscale, a confined lubricant film separating crystalline sliders
undergoes strong layering and solidification.
Under pressure, transitions from $ N \to N-1$ layers may imply a change of lateral periodicity of the
crystallized lubricant sufficient to alter the matching of crystal structures, influencing the ensuing friction jump.
A pressure-induced friction drop may occur as the shear gradient maximum switches from the lubricant
middle, marked by strong stick-slip with or without shear melting, to the crystalline slider-lubricant interface,
characterized by smooth superlubric sliding. We present high pressure sliding simulations to
display examples of frictional drops, suggesting their possible relevance to the local behavior
in boundary lubrication.
\end{abstract}
\date{\today}
\maketitle


\section{Introduction}

Confined lubricants under shear and high pressure display, both experimentally~\cite{israelachvili1,israelachvili2,klein95}
and theoretically~\cite{landman1,landman2,robbins90,perssonbook,braun_rep,vanossi_RMP}, intriguing nano- and meso-scale tribological phenomena.
The intervening lubricant film between two sliding solid surfaces generally changes from liquid (with hydrodynamic lubrication),
to solid or nearly solid at high pressure when the film is only a few monolayers thick.
Both experiments and simulations find that in this regime the film develops a solid-like layered structure,
supporting static friction, and a strong stick-slip frictional behavior.
The stick-slip is often believed to be associated either with a lubricant melting-freezing mechanism, where
the film disorders and melts at slip and solidifies during the stick phase (see \cite{braun_rep} and references therein),
or else with a layer-on-layer shear (a shear band) within the film bulk or at the film-wall interface~\cite{leng2011,klein2007,braun_triblett}.
%
\begin{figure}
\centering
\includegraphics[width=8.5cm,angle=0]{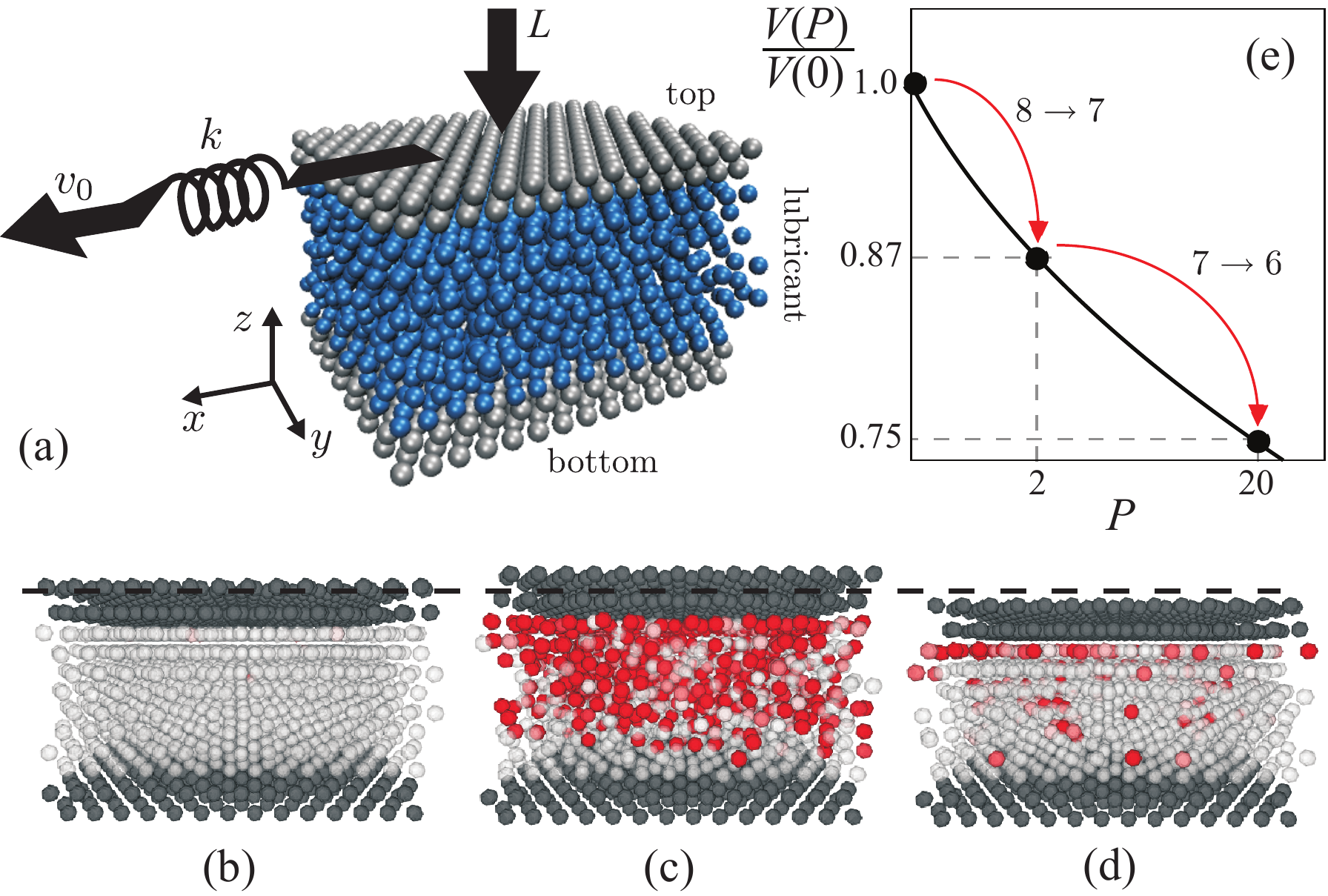}
\caption{(Color online) Sketch of the three-dimensional frictional model (a)
and snapshots of three different sliding states:
(b) 8-layer crystalline lubricant (sticking),
(c) 8-layer melted lubricant (slipping),
(d) 7-layer (relayered) crystalline lubricant, with
a superlubric film-wall interface.
The grey-red colored scale highlights particle kinetic energies, from low to high values.
(e) Bulk lubricant equation of state (sketch) with jumps indicating pressure-induced relayering transitions
under confinement.
}
\label{figura1}
\end{figure}
%
Moreover, as 
lubricant layers get successively squeezed out under increasing pressure,
each $ N \to N-1$ (``relayering'') transition generally corresponds to an upward friction jump~\cite{israelachvili2, landman2}.
Virtually everywhere in sliding friction, including the smoothest boundary lubricated systems, friction grows with load.
The question which we raise here is whether, and under which circumstances, that friction jump could be {\it downward}
rather than upward at relayering. The relative surface structure properties at the interface between sliders and confined film
may play the key role. Under pressure the crystal-like lubricant ordering perpendicular to the interface will typically
be accompanied by ordering parallel to the interface~\cite{robbins_israel,ayappa}.
However, little is known about possible two-dimensional (2D) parallel crystalline order, and even less about its changes
upon high pressure relayering.
There is some evidence, for example in molecular squeezout simulations~\cite{tartaglino} indicating
crystalline parallel ordering of the narrowly confined lubricant, and its dependence on the load.
That detail is important for the case of crystalline sliders, for in general the frictional forces will depend
on the relative geometry and commensurability at the slider-lubricant interfaces, in a way which is not immediately predictable,
in particular not automatically monotonic with pressure.
To explore the variety of possibilities, we decided to abandon the open geometry where lubricant squeezout
takes place, surely much faster with sliding than without~\cite{persson_tosatti, persson_mugele},
and we simulated instead lubricated sliding friction under sealed conditions, with constant particle number and no squeezout.
Sealed conditions might be locally realized in some real cases, but we use them here essentially as a theoretical device.
Our results show that in agreement with pressure-induced $N \to N-1$ relayering of the solidified
confined lubricant film, the parallel lubricant periodicity also changes; and as a result the lubricated sliding friction
may indeed switch from stick-slip, accompanied
by bulk shear with or without melting-freezing, to smooth sliding regimes
typical of incommensurate interfaces exhibiting superlubric dry friction. In that case the friction coefficient
correspondingly shows a strong, if counterintuitive, non-monotonic behavior as a function of increasing load.

\section{Model}

Our model system comprises two parallel, rigid, periodically corrugated plates confining between them
a small number of lubricant layers, with planar periodic boundary conditions (PBC),
with $n$, $n_{t}$ and $n_{b}$ defining the total numbers of particles in the lubricant, top, and bottom plates respectively.
We will present molecular dynamics (MD) simulation results and a detailed statistical data analysis mostly for 2D systems, where the two plates
consist of one-dimensional line-like confining substrates with PBC applied only along the sliding $x$-direction.
We also simulated, not as extensively, more realistic (and computationally more expensive) three-dimensional (3D) systems
such as that depicted in Fig.~\ref{figura1}(a) observing an essentially identical behavior (details in supplemental material).
The bottom plate is assumed to be rigid and immobile while the top sliding plate, whose center of mass coordinate is defined in two dimensions
by $\bf{R}_{\rm top}$ $\equiv$ $(X_{\rm top},Z_{\rm top})$, is connected to an external spring of stiffness $k$ driven
horizontally along $x$ at constant velocity $v_0$.
A downward vertical load force $L$ is applied to the top plate center of mass, $F_N=L/n_t$ being the load per top slider particle.
The corresponding friction force (per top particle) is directly measured by the instantaneous spring elongation
$F=k(X_{\rm top}-v_0 t)/n_t$.
The wall-lubricant (WL) and the lubricant-lubricant (LL) interactions are modeled by Lennard-Jones (LJ) potentials $U_{LL}$ and $U_{WL}$,
choosing an amplitude ratio $\epsilon_{LL}/\epsilon_{WL}$ smaller than $1$ (typically we used $0.2$), so that the lubricant wets the substrate,
favoring epitaxial ordering in the solid phase, such as would generally be the case for example for an octamethylcyclotetrasolixane (OMCTS)
lubricant film between crystalline sliders.
We assume $\sigma_{WW}=1$ as our reference length and define $\sigma_{WL} \equiv (\sigma_{WW}+ \sigma_{LL})/2$.
Acting on the LJ interparticle distance ratio $\sigma_{LL}/\sigma_{WL}$ we can explore the effect of
wall-lubricant interface commensurability changes on the tribological response of the confined system.
The equations of motion for the top plate and the confined lubricant particles are
\begin{align}
\nonumber
M \ddot{\bf{R}}_{\rm top} = &-\sum\limits_{i= 1}^{n_{t}}\sum\limits_{j=1}^{n} \frac{d}{{d {\bf{r}}_{i}^t}}U_{WL}(|{\bf{r}}_{j}-{\bf{r}}_{i}^{t}|)
- k(X_{\rm top}-v_0 t)\hat{{\bf{i}}} \\
&- \sum\limits_{i= 1}^{n}m \eta ({\dot {\bf{R}}}_{\rm top}-{\dot {\bf{r}}}_{i})
- L\hat{{\bf{k}}}+{\bf{F}}^{ran},
\end{align}
\begin{align}
\nonumber
&m \ddot{{\bf{r}}}_i = -\sum\limits_{j \ne i}^{n}\frac{d}{{d {\bf{r}}_{i}}}
U_{LL}(|{\bf{r}}_{i}-{\bf{r}}_{j}|)
-\sum\limits_{j= 1}^{n_{t}}\frac{d}{{d {\bf{r}}_{i}}}
U_{WL}(|{\bf{r}}_{i}-{\bf{r}}_{j}^t|)\\
&+\sum\limits_{j= 1}^{n_{b}}\frac{d}{{d {\bf{r}}_{i}}}
U_{WL}(|{\bf{r}}_{i}-{\bf{r}}_{j}^b|)
- m \eta \dot{{\bf{r}}}_i - m\eta(\dot{{\bf{r}}}_i-\dot{{\bf{R}}}_{top})+{\bf{f}}^{ran}_i,
\end{align}
where ${\bf{r}}_{i}$, ${\bf{r}}_{j}^t$, and ${\bf{r}}_{j}^b$ are the lubricant, top and bottom plate particle coordinates,
respectively, $m$ is the lubricant particle mass, $M=n_{t} m$ that of the top plate, and $\hat{{\bf i}}$ and $\hat{{\bf k}}$
are unit vectors along $x$ and $z$ respectively.
During simulated sliding, the shear-induced Joule heat is removed by means of a viscous damping proportional to the relative velocity
between the top plate and the lubricant, $\eta$ being the friction coefficient.
Importantly here, as detailed later, notwithstanding this rather arbitrary choice of dissipation, picking another thermostating procedure only
influences details, but it does not change the phenomenology essence.
The temperature $T$ is controlled by an ordinary Langevin thermostat, with a random force obeying the fluctuation-dissipation theorem,
i.e. $\langle {\bf{f}}^{ran}_i(t)\rangle=0$ and $\langle {\bf{f}}^{ran}_i(t) {\bf{f}}^{ran}_j(t')\rangle=4 \eta K_B T \delta_{ij}\delta(t-t')$,
$K_B$ being the Boltzmann constant. For the top plate ${\bf{F}}^{ran}=-\sum_{i=1}^{n} {\bf{f}}^{ran}_i$ so that $\langle{\bf{F}}^{ran}(t)\rangle=0$
and $\langle{\bf{F}}^{ran}(t){\bf{F}}^{ran}(t')\rangle=n 4 \eta K_B T \delta(t-t')$.

LJ units are used throughout the paper. The equations of motion are integrated using a modified Velocity-Verlet algorithm
with a sufficiently small time step ($\Delta t=0.005$). Sliding simulations are performed at temperature $K_BT=0.1$.
The external driving velocity is $v_0=0.1$, the spring constant $k/n_t=0.1$, the damping coefficient $\eta=0.2$ (underdamped regime),
and the lubricant particle mass $m=1$. Typical system sizes comprise from hundreds up to a few thousands particles.
Statistical values of the considered physical quantities are obtained by averaging over sufficiently long time intervals
in steady state regimes.
%
\begin{figure}
\centering
\includegraphics[width=8.5cm,angle=0]{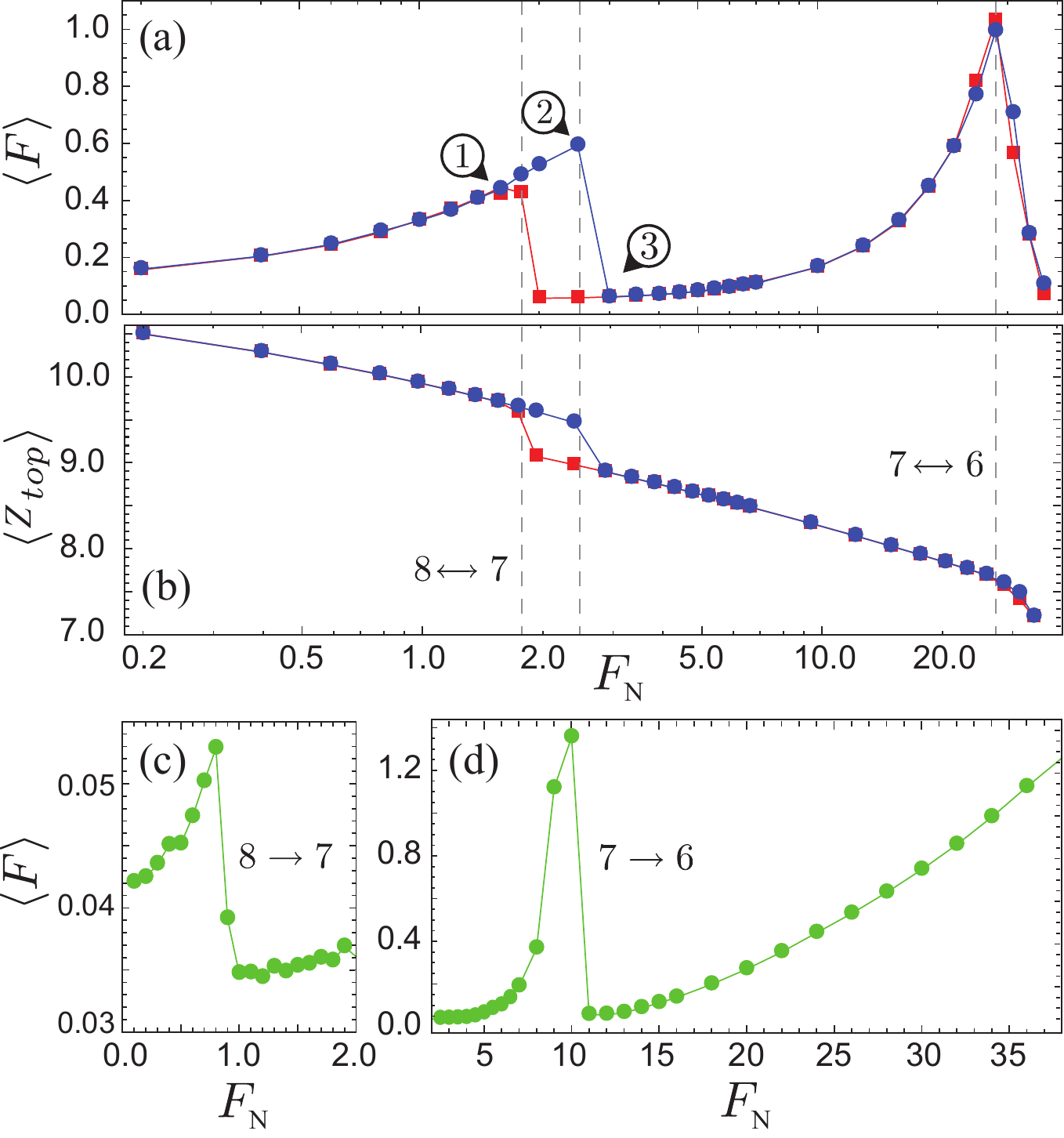}
\caption{(Color online) (a): average friction force versus adiabatically increasing
(circles) and decreasing (squares) vertical load, for the lubricant-slider 2D commensurate system ($\sigma_{LL}/\sigma_{WL}=1$).
A hysteresis loop appears at the relayering transitions of the lubricant.
The numbers $1,2$ and $3$ refer to three different values of load highlighted in Fig.~\ref{figura3}.
(b): load dependence of the lubricant film thickness, measured by the top wall vertical coordinate, $\langle Z_{\rm top} \rangle$.
(c)-(d): average friction force, same as (a) now for 2D lubricant-slider incommensurate system ($\sigma_{LL}/\sigma_{WL} \approx 1.11475$).
The $8 \to 7$ relayering transition here takes place at smaller load, still leading to a friction drop.}
\label{figura2}
\end{figure}

\section{Results}

\subsection{Load-induced drop of friction}

We start off with a film of $N=8$ solid lubricant layers, zero load, and zero sliding speed. Choosing our temperature
above the melting point, the lubricant film first melts. From this point we increase by steps the load $F_N$, thus exerting
a pressure $P$; at first the lubricant recrystallizes with $N=8$ layers. Increasing further the
pressure, the compressed confined lubricant roughly follows (with a kinetics that is fast here given the small system size)
the Lennard-Jones equation of state, as sketched in Fig.~\ref{figura1}(e).
Specifically, the averaged perpendicular interparticle spacing diminishes gradually, as highlighted by the corresponding gentle decrease
of $Z_{top}$ in Fig.~\ref{figura2}(b), up to the occurrence of a relayering transition.
As indicated, the finite system size determines a series of critical pressures $P_{N\rightarrow N-1}$, $P_{N-1 \rightarrow N-2}$, etc,
where relayering takes place, first from 8 to 7 layers, then from 7 to 6, etc, corresponding to significant downward jumps of interparticle
spacing and to a {\it sudden change of the mutual parallel commensurability} between lubricant film and sliders.
The interparticle space shrinking taking place in all directions, at each relayering
the mutual commensurability between lubricant film and sliders undergoes a sudden change.
At this static level, our choice $\frac{\sigma_{WW}}{ \sigma_{LL}} = 1$ and the relatively strong plate-lubricant interaction,
make the plate-lubricant interface initially commensurate in the low pressure state.  That interface becomes incommensurate
as relayering jumps take place under pressure.
When successively the top plate driving speed $v_0$ is raised above zero, and sliding takes place, the frictional force,
and in fact the overall sliding habit, displays a strong dependence upon pressure. While friction naturally increases with load,
there are large jumps at each relayering, where friction now drops, running against conventional wisdom.
%
\begin{figure}
\centering
\includegraphics[width=8.5cm,angle=0]{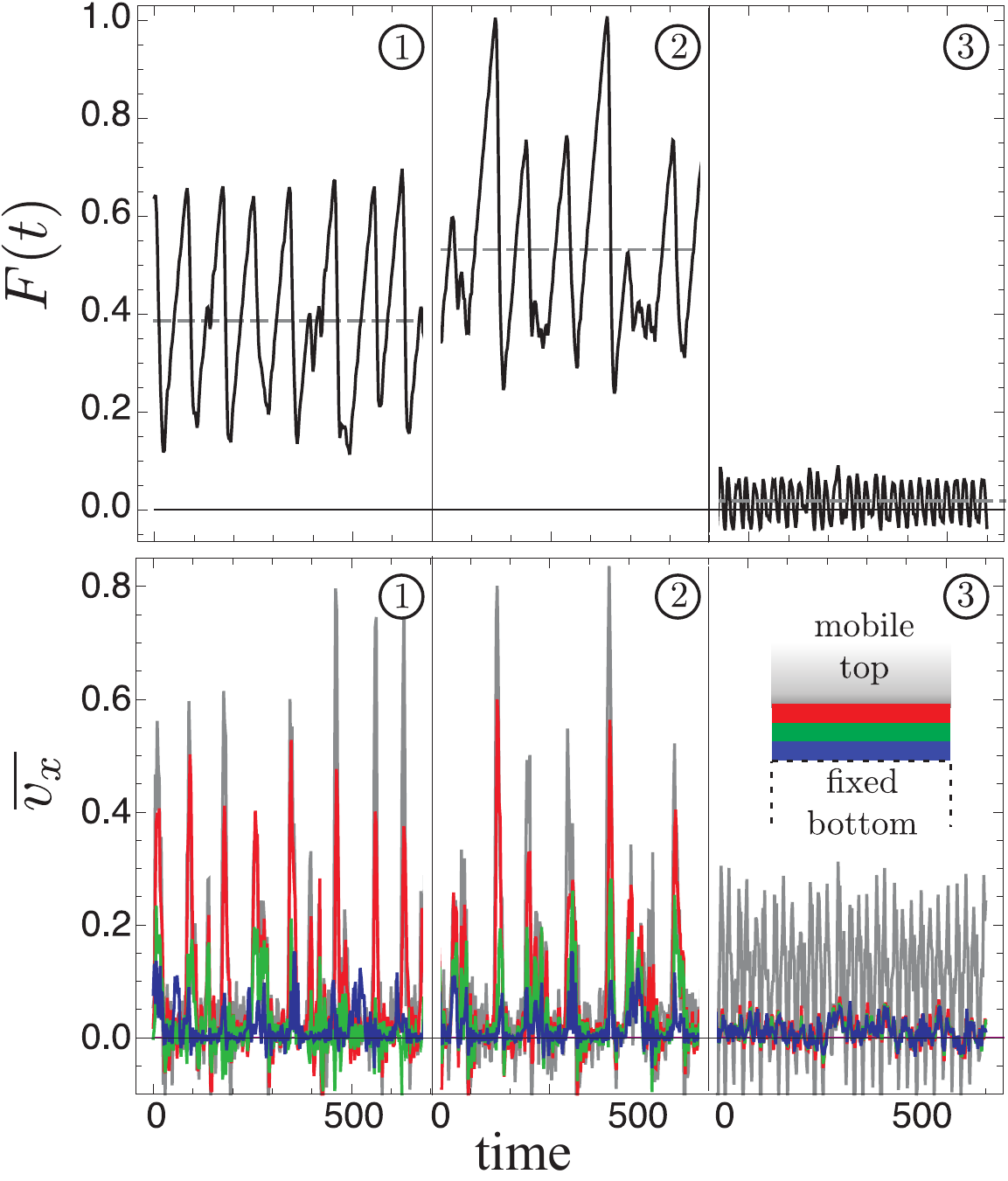}
\caption{(Color online) Upper panels: friction force as a function of time for the three different increasing
loads ($F_N=1.6$, $3.0$ and $3.5$) highlighted in Fig.~\ref{figura2} (a), for the 2D commensurate system
($\sigma_{LL}/\sigma_{WL}=1$). At $F_N=3.5$ (panel $3$) the lubricant film has already undergone the relayering
transition from $8$ to $7$ layers.
Lower panels: corresponding top plate and layer-averaged lubricant velocities, color-coded along $z$ according to the inset sketch.
At the relayering transition sliding switches from the lubricant middle (panels $1$, $2$), shear melting with an
almost laminar flow, to the hard solid wall-lubricant interface (panel $3$).
}
\label{figura3}
\end{figure}
%

Figures~\ref{figura2}(a) and (b) show the non-monotonic pressure behavior of the average friction force $\langle F \rangle$,
and the corresponding average effective film thickness (top wall vertical coordinate) $\langle Z_{\rm top} \rangle$,
for adiabatically increasing (circles) and decreasing (squares) load $F_N$, and an initially commensurate sliding interface
($\sigma_{LL}/\sigma_{WL}=1$).
At small load, with $N=8$ solid lubricant layers, the sliding is smooth and friction is relatively low.
With increasing load, the smooth sliding regime is gradually (although not uneventfully, as will be described later)
replaced by stick-slip. Here friction is much higher, corresponding (for this choice of the interaction parameters)
to shear induced melting-freezing~\cite{robbins90,landman2,perssonbook,braun_rep} originating inside the lubricant,
as visually highlighted in snapshots (b) and (c) of Fig.~\ref{figura1}.
At a critical load $P_{8\rightarrow 7}$, however, the lubricant, unable to support the excessive pressure, undergoes
relayering accompanied by an abrupt, and tribologically crucial, change of parallel lubricant crystalline
periodicity accompanying the drop of perpendicular interlubricant spacing.
In the denser relayered state, two things happen. First, the shear induced melting is now more difficult. Second, and most important,
the interfaces change their mutual commensurability. Thus the relayering causes the shear gradient maximum to switch
from the bulk lubricant middle to the plate-lubricant solid interface (Fig.~\ref{figura1}, snapshot (d)
and movies in the Supplemental Material) whereby stick-slip is replaced by ``superlubric''~\cite{shinjo93,dienwiebel} smooth sliding,
with a dramatic friction decrease. Upon further increase of load, the stick-slip dynamics returns, this time without lubricant melting,
until $P_{7\rightarrow 6}$ and the next relayering take place. Cycling adiabatically the pressure up and down opens up
hysteretic frictional cycles close to relayering transitions (Fig.~\ref{figura2}(a) and (b)).

\subsection{Robustness of friction drop}

This observed frictional drop phenomenon might appear at first strictly determined by our special choice of parameters, and thus not robust;
we found that it is not so.

A variation of the LJ parameter ratio $\sigma_{LL}/\sigma_{WL}$ makes the initial low pressure lubricant-wall interface incommensurate,
yet still leading to clear non-monotonic friction and downward jumps, as shown in Figure~\ref{figura2} (c) and (d).

The same qualitative result is also obtained changing the total number of confined lubricant particles, and by moving to a more realistic 3D model.
%
\begin{figure}
\centering
\includegraphics[width=8.5cm,angle=0]{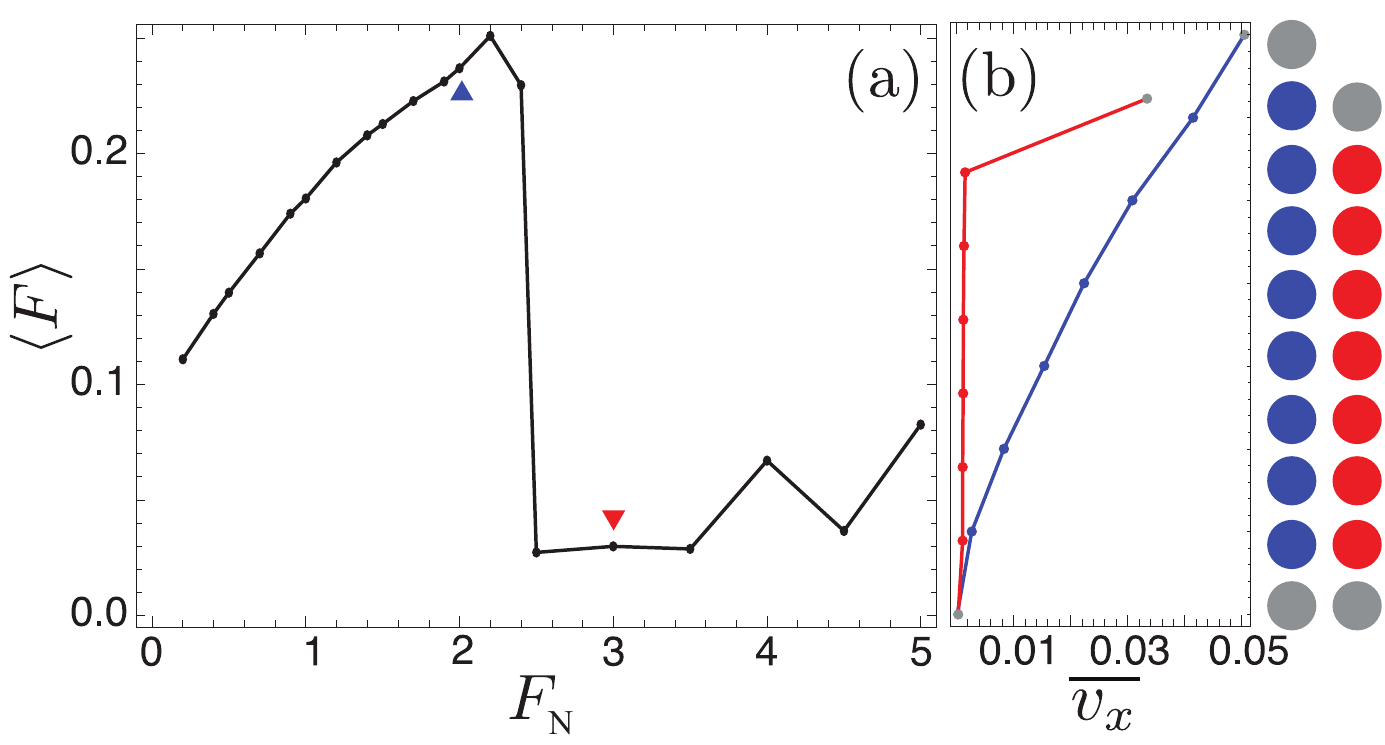}
\caption{(Color online) 3D lubricant-wall system geometry: (a) average friction force versus adiabatically increasing vertical load,
showing a large frictional drop at the $8 \to 7$ relayering transition;
(b) $z$-resolved profile (layer-by-layer) of the lubricant velocity $x$-component.
}
\label{figura4}
\end{figure}
%
In Fig.~\ref{figura4}, the blue data points show an almost laminar lubricant flow of the confined film corresponding
to a high dissipative stick-skip regime of motion. The red points highlight the sliding switch, after pressure-induced relayering,
from the bulk lubricant middle to the incommensurate top wall-film interface, realizing an almost frictionless (superlubric)
dynamical regime (movies in Supplemental Material).

\begin{figure}
\centering
\includegraphics[width=8.5cm,angle=0]{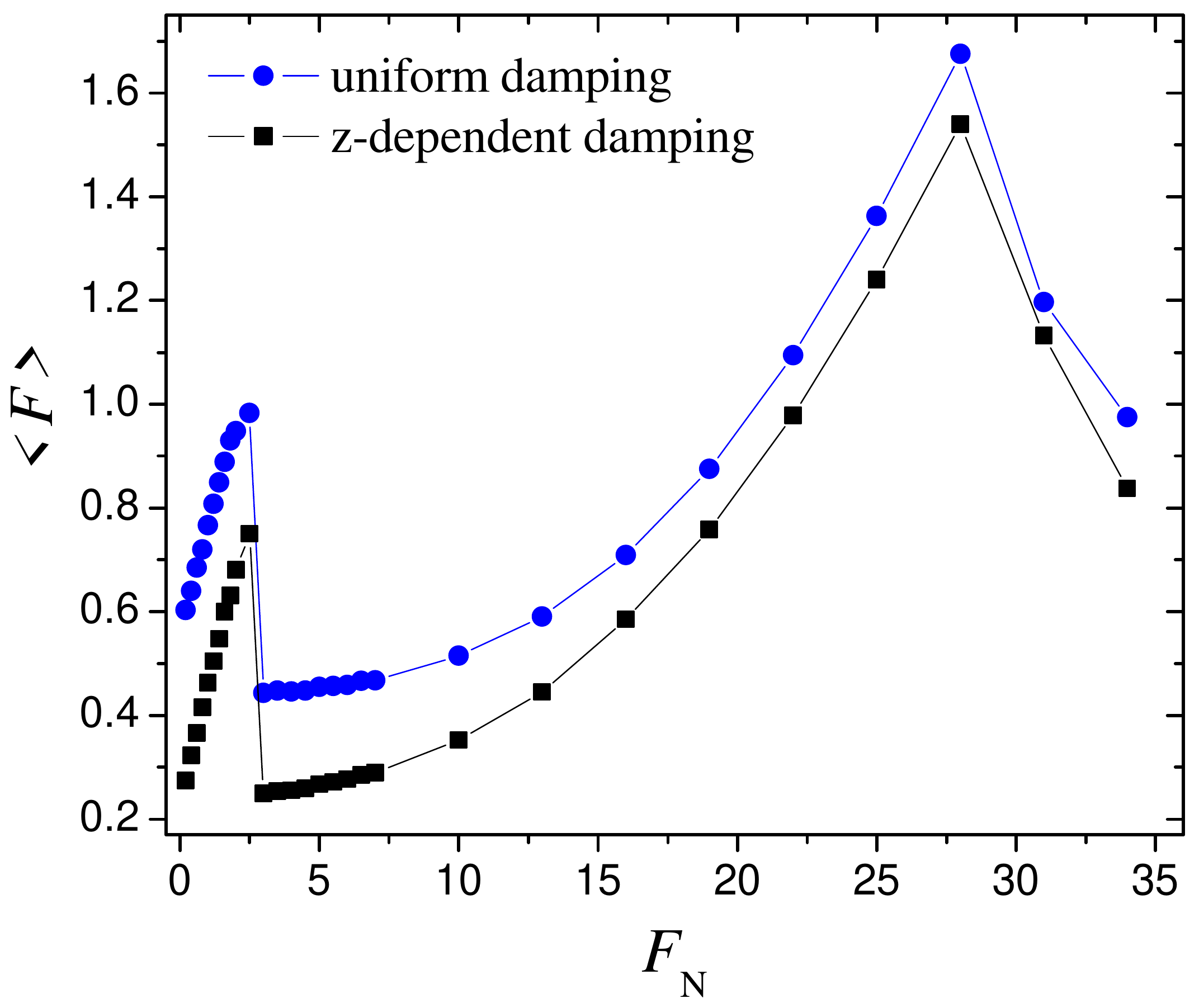}
\caption{(Color online) Average friction force versus adiabatically increasing
vertical load for two distinct dissipation schemes: (circles) a constant and heavy damping
($\eta=2.0$, i.e., $10$ times larger than that used for the results reported in Fig.~\ref{figura2})
on all lubricant particles; (squares) a damping exponentially decaying with distance away from both substrates
(see Ref.~\cite{braun_rep}).
}
\label{figura5}
\end{figure}

Nonequilibrium molecular-dynamics simulations in sliding friction, are also hampered by arbitrariness and uncertainties
in the way Joule heat is removed; in order to attain a frictional steady state, a realistic energy dissipation
is generally impossible to simulate reliably, due to size and time limitations~\cite{benassiPRBrapid,benassi_TribLett}.
However, as shown by the average friction force trend in Fig.~\ref{figura5}, increasing Langevin damping
or else adopting a different thermostat scheme (a damping exponentially decaying with distance away from both sliders~\cite{braun_rep})
only influences details here, without changing the essence: the pressure-induced switches of the maximal shear zone
from the lubricant middle to the slider-lubricant interface, entailing a frictional drop,
is robust and occurs with all thermostating procedures tried.
%
%

Another relevant point strengthening the results' generality deals with the unimportance of whether
dissipative stick-slip shearing through the middle of the solid lubricant film will be accompanied
by melting-freezing, or will just constitute an interlayer shear band.
Experimentally (e.g., in surface force apparatus measurements), one cannot directly observe the film dynamics
and both scenarios easily achievable in MD simulations (depending on the model parameters, geometry and driving conditions)
represent interesting perspectives. However the physics which we describe is not bound to one or the other.
The switch of shear from center to interface which we propose and demonstrate in a particularly simple model,
describes a more general phenomenon, that occurs in our simulations, where forces are rather generic, in either cases.

The time dependent dynamical frictional forces in Fig.~\ref{figura3} show, at loads $F_N$ marked as 1, 2, 3
in Fig.~\ref{figura2}(a), an increasingly strong stick-slip followed by a drop back to smooth sliding
corresponding to a harder solid lubricant with a different parallel commensurability at the relayering transition.
As highlighted by the color-coded lubricant speed along $z$ (lower panels $1,2$, Fig.~\ref{figura3}),
the shear flow profile is almost laminar inside the film, starting off as a shear band at the center of the film.
At relayering (e.g., $8 \to 7$ layers), bulk shear disappears, and the speed gradient switches entirely to the
(now incommensurate, and superlubric) slider-lubricant interface (lower panel $3$, Fig.~\ref{figura3}).
Upon further increase of load, superlubric sliding persists only up to a threshold pressure.
Here, despite incommensurability, static friction and stick-slip sliding reappear, with friction again rising with pressure.
Unlike low pressures, high pressure melting-freezing is suppressed, and stick-slip is ruled by inertia~\cite{braun_rep}.
At each successive relayering transition (eg. from $7 \to 6$ layers), friction may, in principle, jumps down
(as shown in Fig.~\ref{figura2}(a) and (d)) or up depending, respectively, on the realization of a more favorable
or unfavorable interface incommensurability due to lubricant reordering parallel to the confining surfaces.

\begin{figure}
\centering
\includegraphics[width=8.5cm,angle=0]{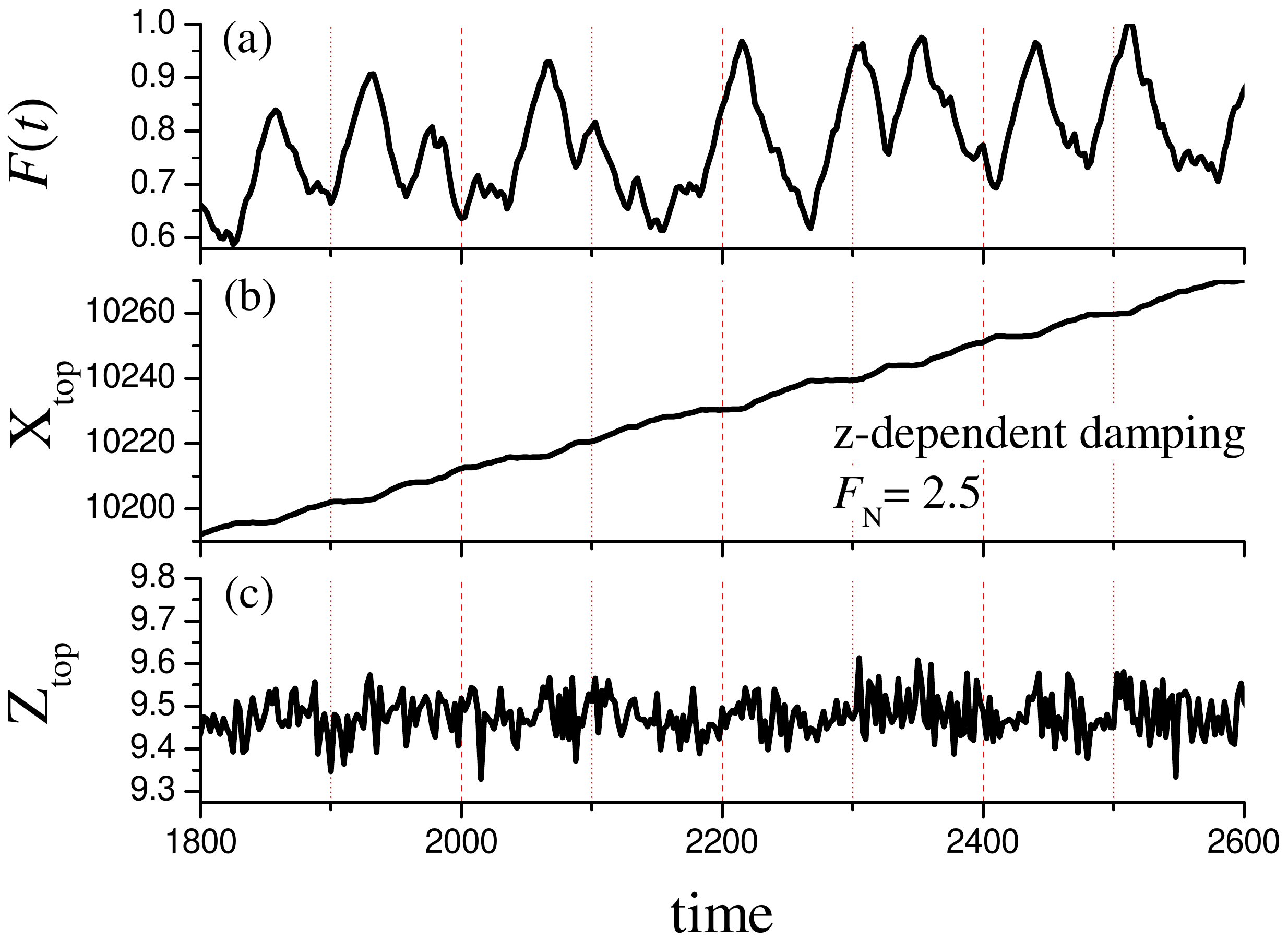}
\caption{Stick-slip dynamics in the presence of a damping exponentially decaying with distance
away from both substrates, at a vertical load $F_N=2.5$ just before the 8 to 7 relayering transition:
(a) frictional force pattern; (b) top wall horizontal $X$-coordinate; (c) top wall vertical $Z$-coordinate.
Here we do not observe a density decrease in mid-lubricant slips, that now leave the film mostly crystalline
and do not lead to shear-induced melting.
}
\label{figura6}
\end{figure}

\begin{figure}
\centering
\includegraphics[width=8.5cm,angle=0]{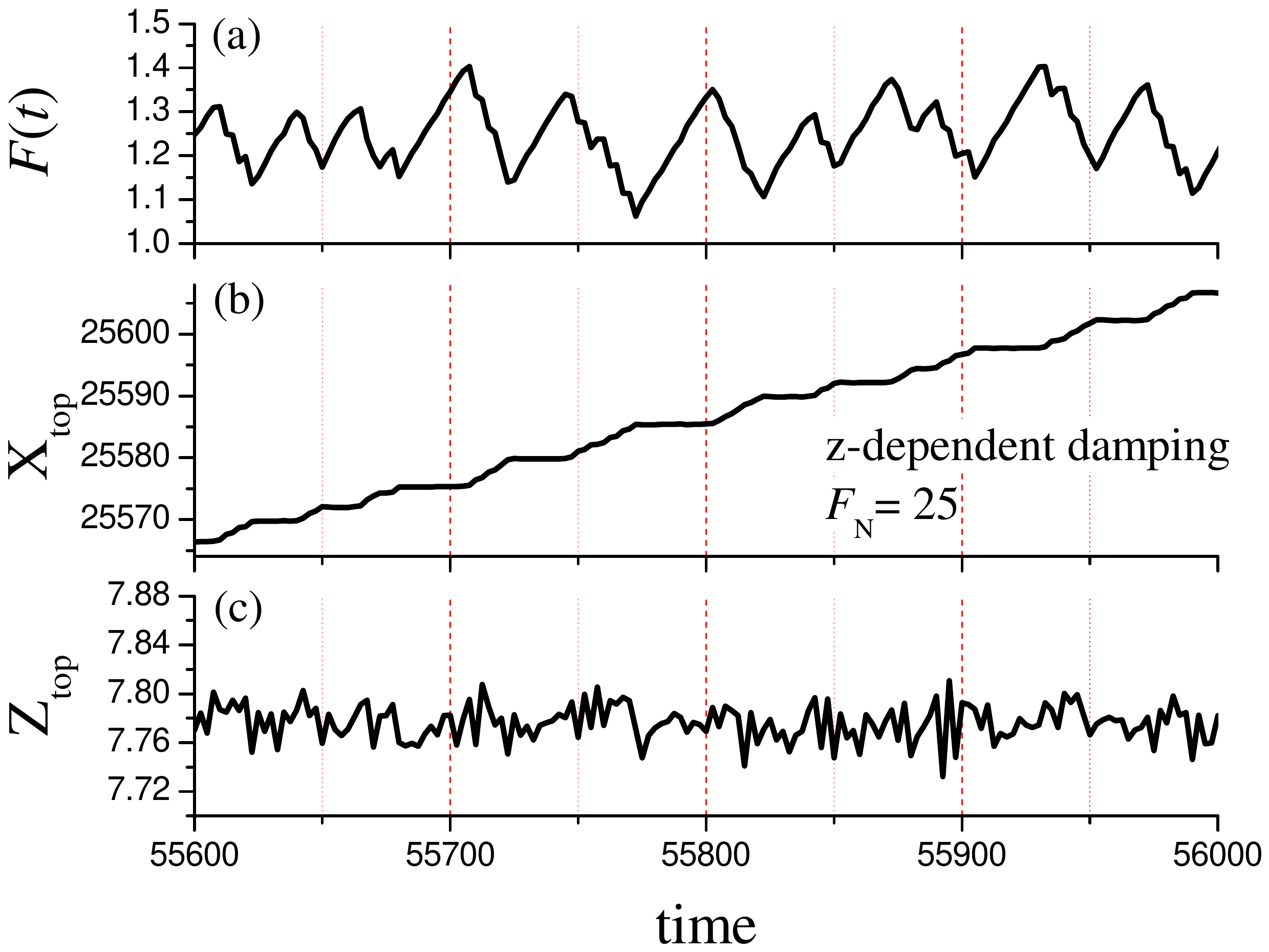}
\caption{Stick-slip dynamics as in Fig.~\ref{figura6} but at a vertical load $F_N=25$ approaching the 7 to 6
relayering transition. Here, apart from the substrate-distance dependent dissipation scheme,
the shear induced lubricant melting during slips is hampered by the strong confining pressure.
}
\label{figura7}
\end{figure}

For the choice of the interaction parameters in Fig.~\ref{figura2}, when melting takes place, the lubricant density decreases.
We do observe that density drop in the form of a small expansion of the film thickness in mid-lubricant slips
(that lead to shear-induced melting), clearly resulting in corresponding vertical jumps of the top confining substrate.
In contrast, the stick-slip dynamics in the presence of a damping exponentially decaying with distance
away from both substrates (square-point curve in Fig.~\ref{figura5}) does not display melting-freezing
induced by sliding, with the top wall vertical $Z_{top}$-coordinate just fluctuating weakly
around a mean value and without any observable density decrease in mid-lubricant slips (panel (c) in Figs.~\ref{figura6} and \ref{figura7}).

%
%

%
%

%
\begin{figure}
\centering
\includegraphics[width=8.5cm,angle=0]{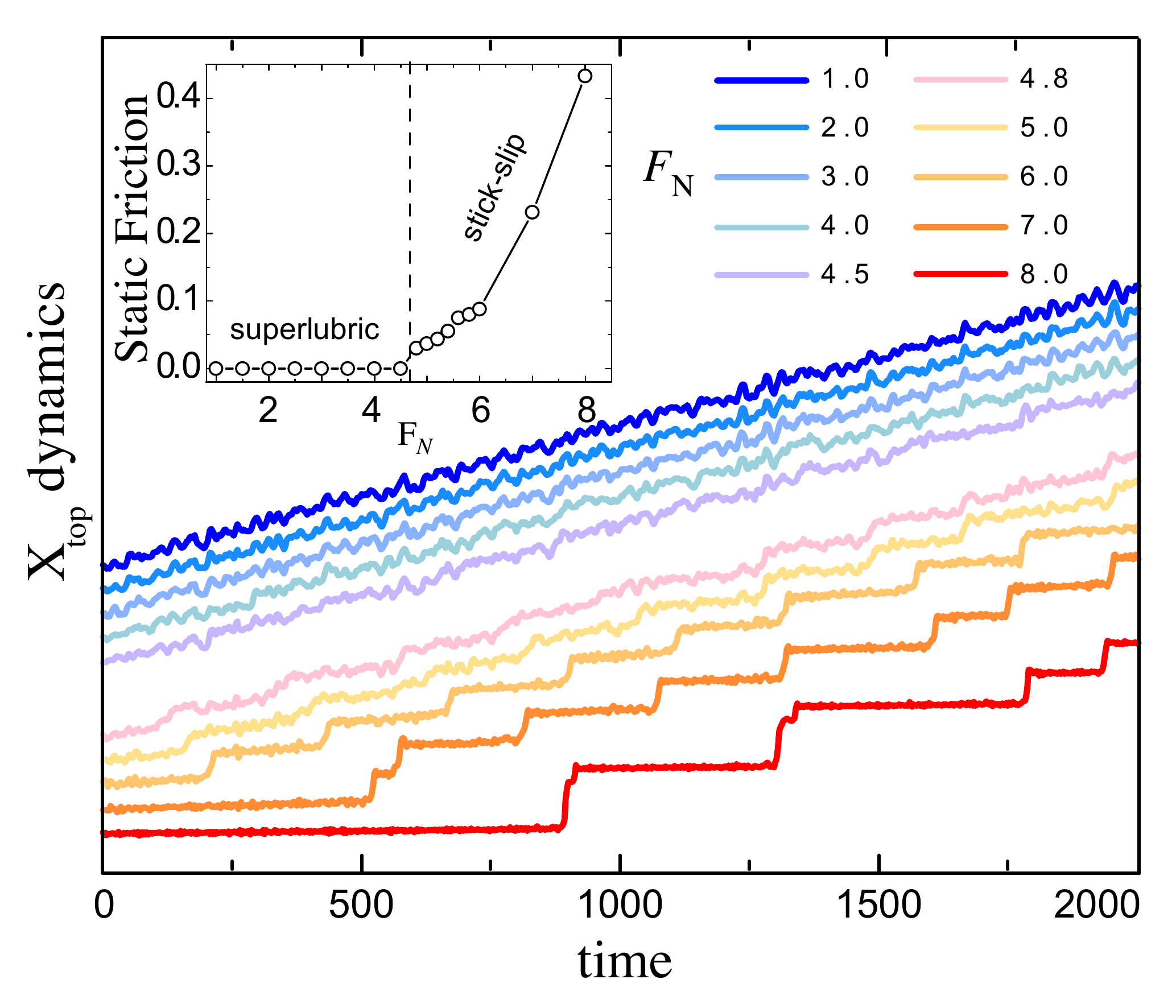}
\caption{(Color online) Time-evolution of the top plate coordinate $X_{\rm top}$
for 2D incommensurate sliding with $\sigma_{LL}/\sigma_{WL} \approx 1.11475$) (same as in Fig.~\ref{figura2}(c)),
for increasing load just after the occurrence of the $8 \to 7$ relayering transition. Curves are offset vertically for clarity.
The inset highlights the Aubry-like load-induced onset of the static friction for the top plate,
marking the transition from superlubric sliding to stick-slip.
}
\label{figura8}
\end{figure}
%

\subsection{Superlubric sliding under pressure}

This pressure induced switch from superlubricity to static friction with stick-slip sliding
deserves special attention, because it takes place continuously and within the same incommensurate
state of the lubricant.
To understand that, we calculated the static friction as follows. Starting from a set of configurations
of the film whose loads are above the $8$ to $7$ relayering transition,
we first removed the external driving and equilibrated the system under the action of increasing loads.
For each configuration,
we evaluated the static friction threshold by driving the top wall external spring
at a very small velocity (i.e., 100 times smaller than that previously adopted)
and measuring the height of the peak preceding the first slip.
Figure~\ref{figura8} shows a clear onset of static friction near $F_N = 4.7$,
taking place without any change in the incommensurate slider-lubricant registry.
This represents an example of the well-known ``Aubry transition''~\cite{peyrard83} in one-dimensional systems,
where the increasing interaction between a harmonic sliding chain and an incommensurate periodic
potential gives rise to dynamic pinning with the onset of static friction.
The sliding trajectories (Fig.~\ref{figura8}) correspondingly switch from smooth sliding
to stick-slip of strength increasing with load.

\subsection{Possible experimental realization}

First let us consider estimates for OMCTS, a reasonably spherical molecule, as a test case lubricant.
Using LJ parameters for OMCTS as provided by H. Matsubara et al.~\cite{matsubara}, the critical pressure
for the $8 \to 7$ relayering is about $6-7$ Kbar. That seems too high compared with ordinary squeezing
loads in surface force apparatus, although local trapping of sealed lubricant ``puddles'' might still
take place at inhomogeneous interfaces~\cite{afm_liq}.

Mesoscopic systems appear more promising. For example, colloids can be confined in layers, crystallized
due to their soft screened Colomb repulsion, and also driven to slide against laser generated periodic
potentials~\cite{bechinger,vanossi_NatMat,vanossi_PNAS} .
They might therefore represent a good trial case for a study relayering of high pressure frictional drops.

\section{Conclusions}

In summary we have explored, by simulating lubricated sliding under sealed conditions, the possibility that
frictional jumps associated with $N \to N-1$ relayering transitions might be downward rather than
upward as commonly observed. We find that downward jumps may occur due to two factors:
(i) high friction with internal shear, easier at low pressure and low density,could disappear after
relayering, when density is higher;
(ii) the change of interface commensurability entails possible superlubricity at higher
pressure. In experiments so far, the friction drop just described is likely removed by lubricant squeezout,
and routes should be considered to overcome that. Despite a clear difficulty in realizing sealed sliding
configurations in ordinary materials and sliding geometries, these two elements stand as results of
general importance, given the relevance of any elements that could control, and especially decrease, friction.
Under the more common open boundary conditions of boundary lubrication, where relayering also
takes place, one still could, depending on conditions and materials, realize one or both points above.
In realistic inhomogeneous conditions, moreover, lubricant ``puddles'' might become sealed under pressure~\cite{afm_liq},
and their sudden yielding could be important for the local sliding dynamics.
Finally, future developments of artificial sliding colloidal systems~\cite{bechinger,vanossi_NatMat} might permit
soon the realization of sealed sliding, and the verification of our results, including, in addition to the downward
jumps, pressure induced Aubry transitions.

\begin{acknowledgments}
This work is part of the Swiss National Science Foundation SINERGIA
Project CRSII2 136287$\backslash 1$. The authors gratefully acknowledge
D. Vanossi for computational resources; and N. Manini and O.M. Braun for discussions.
\end{acknowledgments}


\end{document}